\documentclass[conference]{IEEEtran}
\IEEEoverridecommandlockouts
\usepackage{cite}
\usepackage{amsmath,amssymb,amsfonts}
\usepackage{graphicx}
\usepackage{textcomp}
\usepackage{xcolor}
\usepackage{url}
\usepackage{booktabs}
\usepackage{multirow}
\usepackage{array}
\usepackage{adjustbox}
\usepackage{caption}
\usepackage{algpseudocode}
\usepackage{amsmath}
\usepackage{csquotes}
\usepackage{enumitem}
\usepackage{tikz}
\usetikzlibrary{positioning, shapes, arrows}
\usepackage{array}
\begin{document}
\title{Boli: A dataset for understanding stuttering experience and analyzing stuttered speech}
\author{\IEEEauthorblockN{Ashita Batra\textsuperscript{1}, Mannas Narang\textsuperscript{2}, Neeraj Kumar Sharma\textsuperscript{2}, Pradip K. Das\textsuperscript{1}}
\textsuperscript{1}Department of Computer Science and Engineering, \textsuperscript{2}Mehta Family School of Data Science and AI, 
\\
\textit{Indian Institute of Technology Guwahati, Guwahati-781039, India}\\
{b.ashita@iitg.ac.in, n.mannas@op.iitg.ac.in, neerajs@iitg.ac.in, pkdas@iitg.ac.in}}
\maketitle
\begin{abstract}
There is a growing need for diverse, high-quality stuttered speech data, particularly in the context of Indian languages. This paper introduces Project Boli, a multi-lingual stuttered speech dataset designed to advance scientific understanding and technology development for individuals who stutter, particularly in India. The dataset constitutes (a) anonymized metadata (gender, age, country, mother tongue) and responses to a questionnaire about how stuttering affects their daily lives, (b) captures both read speech (using the Rainbow Passage) and spontaneous speech (through image description tasks) for each participant and (c) includes detailed annotations of five stutter types: blocks, prolongations, interjections, sound repetitions and word repetitions.
We present a comprehensive analysis of the dataset, including the data collection procedure, experience summarization of people who stutter, severity assessment of stuttering events and technical validation of the collected data.
The dataset is released as an open access to further speech technology development.
\end{abstract}
\begin{IEEEkeywords}
Indian stuttered speech dataset, Intelligibility assessment, read speech, spontaneous speech, Stuttering event detection
\end{IEEEkeywords}
\section{Introduction}

\begin{figure*}[htpb]
 \centering
 \includegraphics[width=1\linewidth]{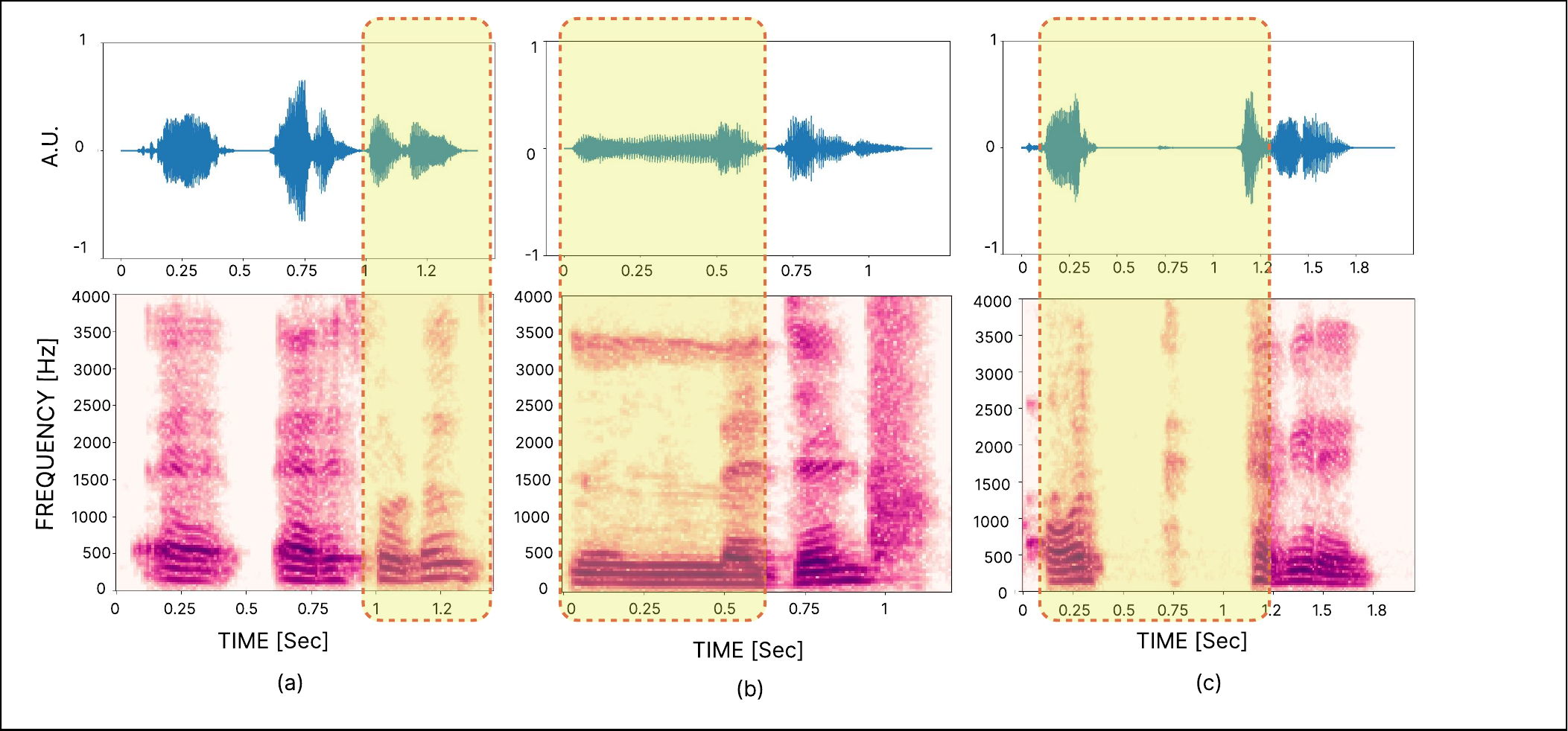} 
 \vspace{-0.3cm}
 \caption{Spectrograms illustrating a few stutter types associated with a male speaker: (a) Sound Repetition (SR), (b) Prolongation (PR), and (c) Block (B).}
 \label{fig:audio_spectrogram_withlabels}
\end{figure*}

\noindent Stuttering (or stammering) refers to atypical speech patterns characterized by significant uncontrollable pauses, filler words (e.g. \textit{umm}, \textit{uhh}, syllable or word repetitions, and other speech disturbances. Table \ref{tab:stutter_type} details these atypicalities, while Figure \ref{fig:audio_spectrogram_withlabels} presents spectrographic visualizations of some examples.

Stuttering affects approximately 12 million individuals in India \cite{b2}. Despite the recent proliferation of voice-based AI assistants like Cortana, Siri, Alexa and Google Assistant, these technologies are significantly challenged when processing and recognizing stuttered speech \cite{b5,b6,b11,b12}. While several datasets have been created in this field, there remains a lack of balanced classes, word-level transcriptions and speaker information related to stuttering experiences.

Existing research on stutter analysis focuses primarily on the following datasets (tabulated in \ref{tab:stutter_datasets}):
\textbf{UCLASS \cite{b13}}: A publicly available clinical English dataset from the University College of London (UCL), focusing on stuttered speech in children aged 7-17 years. It comprises 1 hour of audio recordings, with 138 recordings in release 1, containing annotations for 25 children (2 female, 23 male). Studies using this dataset have employed spectrograms as features and trained Bi-LSTMs for multi-class classification of different stutter types \cite{b19}. This work was extended in \cite{b16}, replacing residual layers with squeeze \& excitation (SE) layers and an attention mechanism.
\textbf{KSoF \cite{b14}}: A German speech therapy-based stuttered speech dataset containing 214 recordings from 37 individuals.
\textbf{Sep-28k \cite{b15}}: The largest publicly available English stutter dataset, comprising 28,177 audio files ($\approx3$ seconds each) from natural conversations in podcasts. It provides file-level annotations for various stutter types: prolongation (PR), interjection (IN), sound repetition (SR), word repetition (WR) and block (B).
\textbf{FluencyBank \cite{b18}}: An interview-based dataset of 32 individuals ($\approx3.5$ hours) with labeling similar to the Sep-28k dataset.
Recent additions to stutter datasets include a Mandarin corpus \cite{b30}, which is twice the size of Sep-28k, and a syllable-level stutter dataset in Kannada \cite{b31}. Additionally, \textbf{LibriStutter \cite{b16}} is a publicly available, synthetically generated English dataset derived from the LibriSpeech ASR corpus, containing time-aligned transcriptions from 20 hours of audio data (50 individuals: 23 male, 27 female).

This paper introduces the $Boli$ dataset, documenting demographic information, experiential data, and read and spontaneous speech recordings from individuals who stutter. The Hindi word $Boli$ means \textit{ a person's unique speaking style} and hence is considered an appropriate name for this dataset. The dataset was collected through crowd-sourcing\footnote{The project website used for collecting the data can be accessed through \url{https://project-boli.vercel.app/}} and is manually curated and validated. Section~\ref{sec:data_collection} documents the data collection and curation procedures. Section~\ref{sec:validation} presents the validation of the dataset for the detection of stuttering from speech recordings. Section\ref{sec:questionnaire_summary} presents a summary of the experiential data collected using a questionnaire. Section \ref{sec:conclusion} presents the conclusion.




\begin{table}[htpb]
    \centering
    \caption{Types of Stutter}
    \huge
    \resizebox{\columnwidth}{!}{    
    \begin{tabular}{l|l}
    \hline
    \textbf{Stuttering Type} & \textbf{ Definition \& Examples} \\
    \hline
Prolongation (PR) & Extended phonemes within a word. Eg: {\itshape "S[sss]ee."} \\
Block (B) & Pauses or stoppages in speech. Eg: {\itshape "My \ldots name."} \\
Sound Repetition (SR) & Repetition of phonemes or syllables. Eg: {\itshape "[bu-bu-]butterfly."} \\
Word Repetition (WR) & Repetition of whole words. Eg: {\itshape "I [I] am."} \\
Interjection (IN) & Insertion of unnecessary phonemes or syllables. Eg: {\itshape "[Um, uh,] yes!"} \\
    \hline
    \end{tabular}
    }
    \label{tab:stutter_type}
\end{table}

\begin{table*}[h!]
\centering
\resizebox{\textwidth}{!}{
\begin{tabular}{l c c c c c c}
\hline
\textbf{Dataset} & \textbf{Duration (hours)} & \textbf{Speakers} & \textbf{Transcription} & \textbf{Tasks} & \textbf{Language} & \textbf{Questionnaire}\\ \hline

LibriStutter \cite{b19} & 20 & 50 & Yes & audiobook & English & No \\ 
UCLASS\cite{b13} & 0.88 & 25& Yes & conversation & English  & No \\ 
SEP-28k \cite{b15} & 23& not reported & No & podcast & English  & No \\ 
FluencyBank \cite{b18} & 3.5 & 32 & Yes & conversation and reading & English  & No \\ 
KSoF \cite{b14} & 4.6 & 37  & No & spontaneous and reading & German  & No \\ 
AS-70 \cite{b30} & 48.8 & 72 & Yes & conversation, voice commands & Mandarin  & No  \\ 
Kannada Stutter Dataset \cite{b31}  & 1.4 & 80 & Yes & read \& spontaneous & Kannada  & No \\ 
\textbf{Boli (Proposed)} &\textbf{ 2.5} & \textbf{28}  & \textbf{ Yes} & \textbf{read \& spontaneous} & \textbf{English, Hindi, Telugu, Bengali, Marathi, Assamese}  & \textbf{Yes} \\ \hline
\end{tabular}
}
\caption{Comparison of stuttered speech datasets. Here, PWS stands for people who stutter and CWS for children who stutter.}
\label{tab:stutter_datasets}
\end{table*}

\begin{table*}[h!]
    \centering
    \caption{ Description on metadata relating to stuttered speech utterances}
    \begin{tabular}{lcccc}
        \toprule
        \textbf{Variable} & \textbf{Mean} & \textbf{Standard Deviation} & \textbf{Min} & \textbf{Max} \\
        \midrule
        Age & 26.23 & 7.26 & 17 & 48 \\
        \midrule
        \multicolumn{5}{c}{\textbf{Counts by Category}} \\
        \midrule
        Gender (Male/Female) & \multicolumn{4}{c}{25 / 3} \\
        Stuttering Severity (Mild/Moderate/Severe) & \multicolumn{4}{c}{9/14/5} \\
        \bottomrule
    \end{tabular}
    \label{table:demographics}
\end{table*}
\begin{table}[htpb]
    \centering
    \caption{Average Stuttering Rate for Read \& Spontaneous speech}
    \resizebox{\columnwidth}{!}{    
 \begin{tabular}{lcc}
        \toprule
        \textbf{Speech Type} & \textbf{Avg.Stuttering events/minute} & \textbf{Total stuttering events} \\
        \midrule
        \textbf{Read} & 6.76  & 220 \\
        \textbf{Spontaneous} & 2.02 & 60\\
        \bottomrule
    \end{tabular}
    }
    \vspace{0.1cm}    
    \label{asr}
\end{table}
\section{Data Collection and Curation}
\label{sec:data_collection}
\noindent Data are collected from participants in a crowd-sourcing manner, through a custom-designed website for this task. No person identification data is collected. A participant first fills the demographic information (namely, age, gender, country, city and mother tongue), and then completes a questionnaire on experiential information in relation to stuttering. The questionnaire is composed of $25$ multiple choice questions (MCQs). Subsequently, the participant's speech is collected as audio files corresponding to a few (fixed) sentences from the widely used $Rainbow$ passage (in English and also translated to their respective mother tongues). As a follow-up to this recording, the participant also describes an image (in English and their mother tongue) and this is stored as spontaneous speech audio recording. Participants were encouraged to relax, take time and not produce artificial stutter.
In total, as of 05-Sep-2024, $67$ individuals who stutter (in the age group $17-48$ years) participated in the data collection.

\noindent Manual annotation of stuttered speech recordings is a laborious task due to variations in speaking style and duration across stutter types, necessitating repeated listening of audio files. Consequently, most stuttered speech datasets provide only file-level stutter-type annotations. In contrast, the Boli dataset offers word-level annotations, including specific timestamps and stutter types.
After manual analysis of $67$ participants data, $28$ were identified as containing stutter ($25$ male and $3$ female). The target languages were English and the participants' mother tongues, distributed as follows: Hindi (22), Telugu (2), Bengali (2), Marathi (1), and Assamese (1). Each participant recorded on average $4-7$ minutes, producing approximately $2.8$ hours of audio data. Five types of stutter (see Table \ref{tab:stutter_type}) were identified, with the following distribution of occurrences: Sound Repetition ($SR=140$), Block ($B=70$), Prolongation ($PR=41$), Word Repetition ($WR=21$), and Interjection ($IN=8$). Following are the observations post listening and manual annotations of all the audio recordings.
\begin{itemize}
\item Description on metadata relating to stuttered speech utterances is provided in Table~ref{table:demographics}.
\item Stuttering was significantly less frequent in spontaneous speech compared to read speech. Table \ref{asr} depicts the \textit{Average Stuttering Rate} (number of stuttering events per minute) for both read and spontaneous speech, across participants.
\item Some speakers with severe stuttering produced longer duration samples compared to others.
\item A few participants who self-reported moderate stuttering severity showed no stuttering during recording.
\item Phonemes such as $/r/$, $/t/$, $/p/$, $/f/$, $/sh/$, $/th/$ and words such as $hebrews$, $boiling$, $pot$, $beyond$, $his$ were identified as common stuttering triggers for some speakers.
\end{itemize}


\begin{figure}[htpb]
\centering
\begin{tikzpicture}[
    node distance=1.5cm,
    every node/.style={rectangle, draw, rounded corners, align=center},
    arrow/.style={->, thick}
]
\node (start) [align=center] { Stuttered Speech Signal};
\node (preprocessing) [below of=start] {\textbf{Data Pre-processing} \\ Amplification \\ Normalization};
\node (feature) [below of=preprocessing, node distance=2 cm] {\textbf{Feature Extraction}\\ MFCC};
\node (classification) [below of=feature] {\textbf{Classification Models} \\RF/SVM/LSTM/BiLSTM}; give a caption
\node (pr) [below of=classification] {PR};
\node (b) [left of=pr, xshift=+0.2cm] {B};
\node (in) [left of=pr, xshift=-1.3cm] {IN};
\node (sr) [right of=pr, xshift=0.2cm] {SR};
\node (wr) [right of=pr, xshift=2cm] {WR};
\draw[arrow] (start) -- (preprocessing);
\draw[arrow] (preprocessing) -- (feature);
\draw[arrow] (feature) -- (classification);
\draw[arrow] (classification) -- (pr);
\draw[arrow] (classification) -- (b);
\draw[arrow] (classification) -- (in);
\draw[arrow] (classification) -- (sr);
\draw[arrow] (classification) -- (wr);
\end{tikzpicture}
\caption{Proposed methodology for stutter-type classification from stuttered speech signals.}
\label{flowchart}
\end{figure}
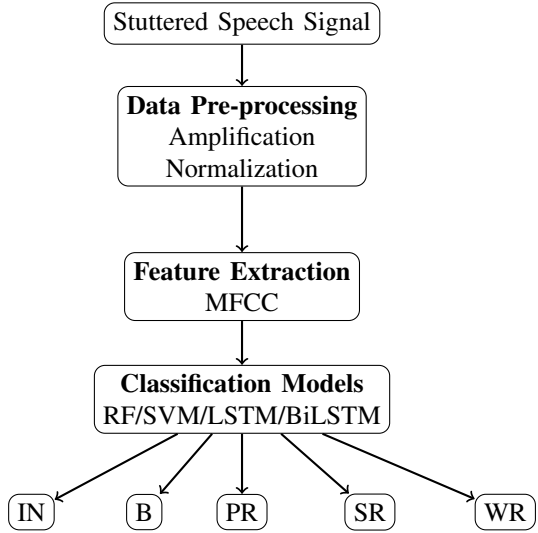



\begin{table*}[htpb]
    \centering
    \caption{Comparison of F1-scores for stuttering event detection using RF, SVM, LSTM and BiLSTM with balanced (B) and imbalanced (ImB) dataset, trained on Sep-28k and testing on Boli dataset, using MFCC features}
    \begin{adjustbox}{max size={\textwidth}{\textheight},center}
    \renewcommand{\arraystretch}{1.5} 
    \setlength{\tabcolsep}{4pt} 
    \begin{tabular}{lccccccccccccccc}
        \toprule
        \textbf{\textbf{Dataset}} & \textbf{\textbf{Method}} & \multicolumn{2}{c}{\textbf{\textbf{PR}}} & \multicolumn{2}{c}{\textbf{\textbf{B}}} & \multicolumn{2}{c}{\textbf{\textbf{SR}}} & \multicolumn{2}{c}{\textbf{\textbf{WR}}} & \multicolumn{2}{c}{\textbf{\textbf{IN}}} & \multicolumn{2}{c}{\textbf{\textbf{Avg. F1}}} \\
        \cmidrule(lr){3-4} \cmidrule(lr){5-6} \cmidrule(lr){7-8} \cmidrule(lr){9-10} \cmidrule(lr){11-12} \cmidrule(lr){13-14}
        & & \textbf{\textbf{B}} & \textbf{\textbf{ImB}} & \textbf{\textbf{B}} & \textbf{\textbf{ImB}} & \textbf{\textbf{B}} & \textbf{\textbf{ImB}} & \textbf{\textbf{B}} & \textbf{\textbf{ImB}} & \textbf{\textbf{B}} & \textbf{\textbf{ImB}} & \textbf{\textbf{B}} & \textbf{\textbf{ImB}} \\
        \midrule
        \multirow{4}{*}{\textbf{\textbf{Boli}}} 
         & \textbf{\textbf{RF}} &\textbf{ 0.93} &  \textbf{0.66} & \textbf{0.76} & \textbf{0.67} & \textbf{0.75} & \textbf{0.78} & \textbf{0.94} & \textbf{0.32} & \textbf{0.99} & \textbf{0.71} & \textbf{0.87} & \textbf{0.71}\\
         & \textbf{\textbf{SVM}} & 0.70 & 0.47 & 0.51 & 0.48 & 0.50 & 0.72 & 0.64 & 0.20 & 0.88 & 0.51 & 0.65 & 0.61 \\
         & \textbf{\textbf{LSTM}} & 0.62 & 0.35 & 0.38 & 0.43 & 0.44 & 0.69 & 0.55 &  0.10 & 0.88 & 0.38 & 0.58 & 0.56 \\
         & \textbf{\textbf{BiLSTM}} & 0.66 & 0.40 & 0.43 & 0.48 & 0.45 & 0.70 & 0.60  & 0.10 & 0.91 & 0.49 & 0.61 & 0.58 \\
        \bottomrule
    \end{tabular}
    \end{adjustbox}
    \vspace{0.2cm}   
    \label{results}
\end{table*}

    

    

\begin{table}[htpb]
    \centering
    \caption{Word Error Rate (WER) performance for: concatenated speech (pooling utterances from all participants) and speaker-wise speech}
    \begin{tabular}{lcc}
        \toprule
        \textbf{ASR Model} & \textbf{Concatenated Speech} & \textbf{Speakers-wise Speech} \\
        \midrule
        \textbf{Wav2Vec2.0} & 29.22\% & 55.81\% \\
        \textbf{Whisper} & 4.55\% & 21.27\% \\
        \bottomrule
    \end{tabular}    
    \vspace{0.1cm}    
    \label{wer}
\end{table}

\section{Stutter-type Classification}
\label{sec:validation}
\noindent To validate the utility of the audio dataset, we analyzed the performance of stutter-type classification by applying a variety of machine learning models on its audio files. We do not aim to achieve state-of-the-art results. Instead, we show the effect of class imbalance (and balancing) and performance comparison amongst traditional machine learning and deep learning-based classification models.
Each audio file had a sampling rate of $16$~kHz. As features, we used the mel-frequency cepstral coefficients (MFCCs). Specifically, $13$ dimensional MFCCs were extracted which were averaged across frames and then fed as input vector, with $25$~ms temporal window and $10$~ms hop duration, resulting in $\approx465$ frames per file. We use the machine learning models: random forest (RF), support vector machine (SVM), LSTM, and BiLSTM. The classification methodology is shown in Figure \ref{flowchart}. Owing to the small size of the Boli dataset, we have focused on performing cross-dataset evaluation, that is, training on Sep-28k and testing on the Boli dataset (English utterances only). The results are shown in Table \ref{results}.
Our experiments maintain consistency in class types and duration ($\approx 5$~ s) to help the model learn characteristics more efficiently. As the Sep-28k dataset is imbalanced in relation to stutter-types, we have used hybrid sampling before training to balance the classes by combining up-sampling and down-sampling. This takes the average of minority and majority samples, computed by dividing these samples by two. It is computed as follows:
\begin{equation}
N_{avg} = \frac{N_{1} + N_{2}}{2}
\end{equation}
where \( N_1 \) is the number of samples in the minority class and \( N_2 \) is the number of samples in the majority class. Balancing classes prevents the model bias towards the majority class, as shown in Table \ref{results}. Comparative analysis between an imbalanced and balanced dataset has been demonstrated using different learning techniques, namely SVM, RF, LSTM and BiLSTM. Table \ref{results} indicates that the RF best captures the characteristics of all stutter classes. Furthermore, statistical technique, i.e., F1-score, is used to assess the model's effectiveness.


In another analysis, we compared two ASR models, namely Wav2Vec2.0 \cite{b27} and Whisper \cite{b28}, to assess their effectiveness in capturing stuttered speech by calculating the word error rate (WER) for English audio recordings. 
We evaluated ASR for two settings: concatenated speech (grouping utterances from all participants), ASR evaluation, and speaker-wise speech evaluation (averaging the performance obtained across participants).  The results are shown in Table \ref{wer}.
On manually verifying the ASR transcriptions, it was found that word repetition (WR) stutter type was well identified by both models. 
\begin{figure}
    \centering
    \includegraphics[width=0.7\linewidth]{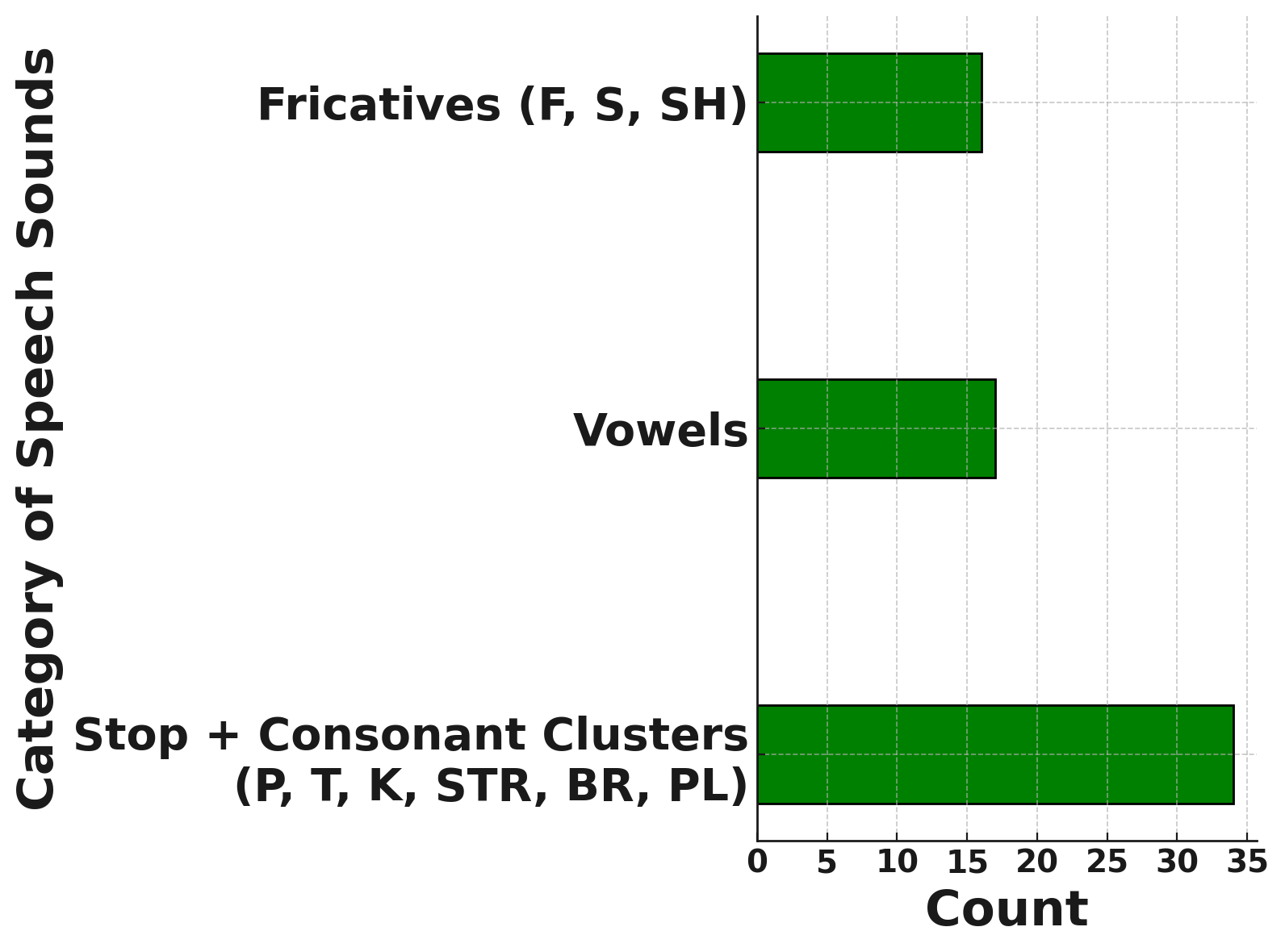}
    \caption{Most stuttered sounds based on information shared through the questionnaire form (collected from 67 subjects)}
    \label{sounds}
\end{figure}
\section{Questionnaire Data}
\label{sec:questionnaire_summary}
From the questionnaire responses of 67 participants, it was observed that language plays a crucial role in stuttering. Most participants reported difficulties with stop consonants (and stop consonant clusters or blends) such as $/p/$, $/t/$, $/k/$,$/str/$,$/br/$ and $/pl/$ (see Figure \ref{sounds}). Many use speech therapy techniques to manage their stuttering and experience similar levels of stuttering in their mother tongue. In general, most have undergone speech therapy and feel hesitant to speak in front of crowds, often experiencing facial muscle tension. Stuttering is more pronounced when speaking to unfamiliar people, leading some to either minimize their stutter or avoid the situation, while only a few disclose it upfront. Singing, however, tends to reduce stuttering, which participants attribute to the rhythm and timing of music, as well as different breathing techniques that help alleviate pressure and anxiety. Additionally, many report being able to anticipate when they might stutter.

\section{Conclusion}
\label{sec:conclusion}
\noindent We present the Boli data set to facilitate the development of technology for people who stutter. It includes both read and spontaneous speech with manual word-level annotations. Every file contains a single stuttered word along with a few non-stuttered words. Additionally, we collected individuals who stutter life experiences through a questionnaire to cross-validate severity, intelligibility, and identifying common stuttered words, syllables across individuals. Despite its small size, the data set includes various types of stutter in both read and spontaneous modes, which is rare in the Indian speech context. This can help improve the performance of the ASR model in multiple languages when the data set expands.


 \end{document}